\documentclass[11pt]{article}

\usepackage[utf8]{inputenc}

\usepackage{amsmath,amssymb}

\usepackage{graphicx}
\graphicspath{{./}}

\usepackage{booktabs}
\usepackage{multirow}
\usepackage{caption}
\usepackage{subcaption}

\usepackage{placeins}   
\usepackage{url}
\usepackage{xcolor}
\usepackage{comment}

\newenvironment{frontmatter}{}{}
\newenvironment{backmatter}{}{}

\providecommand{\chaptermark}[1]{}
\providecommand{\IntechOpentext}[1]{}

\providecommand{\keywords}[1]{\textbf{Keywords:} #1}

\providecommand{\chapter}[1]{\section{#1}}

\title{Security and Resilience in Autonomous Vehicles: A Proactive Design Approach}
\author{Chieh Tsai \and Murad Mehrab Abrar \and Salim Hariri}
\date{}

\begin{document}
\maketitle
\noindent\textbf{Accepted for publication in:
Connected, Cooperative and Autonomous Mobility
(IntechOpen)}


\begin{frontmatter}


\begin{abstract}
Autonomous vehicles (AVs) promise efficient, clean and cost-effective transportation systems, but their reliance on sensors, wireless communications, and decision-making systems makes them vulnerable to cyberattacks and physical threats. This chapter presents novel design techniques to strengthen the security and resilience of AVs. We first provide a taxonomy of potential attacks across different architectural layers, from perception and control manipulation to Vehicle-to-Any (V2X) communication exploits and software supply chain compromises. Building on this analysis, we present an AV Resilient architecture that integrates redundancy, diversity, and adaptive reconfiguration strategies, supported by anomaly- and hash-based intrusion detection techniques. Experimental validation on the Quanser QCar platform demonstrates the effectiveness of these methods in detecting depth camera blinding attacks and software tampering of perception modules. The results highlight how fast anomaly detection combined with fallback and backup mechanisms ensures operational continuity, even under adversarial conditions. By linking layered threat modeling with practical defense implementations, this work advances AV resilience strategies for safer and more trustworthy autonomous vehicles.
\end{abstract}


\end{frontmatter}


\section{Introduction}
The emergence of Autonomous Vehicles (AVs) represents a transformative shift in transportation, with the potential to drastically improve road safety, traffic efficiency, and accessibility to mobility ~\cite{litman2017autonomous, shladover2018connected}. By integrating complex subsystems such as perception, localization, planning, and control, AVs operate with minimal or no human intervention. These systems rely on a wide range of sensors (e.g., LiDAR, cameras, radar), communication protocols (e.g., V2V, V2I), and artificial intelligence (AI) algorithms to make decisions in real time. However, this increasing system complexity also introduces new vulnerabilities~\cite{petit2014potential, checkoway2011comprehensive}. As AVs become more connected and data-driven, their attack surface expands, exposing them to a wide range of cyber and physical threats that can compromise safety, reliability, and user trust.

Recent incidents and research have highlighted that autonomous systems are susceptible to adversarial manipulation at multiple layers—from sensor spoofing and GPS interference to model poisoning and control signal injection\cite{cao2019adversarial, shepard2012drone}. These attacks not only pose risks to the integrity and availability of autonomous functions but may also lead to catastrophic physical consequences, including loss of life collisions and traffic disruption. Therefore, ensuring the security and resilience of AVs have become a critical research challenge, requiring robust threat modeling, proactive defense mechanisms, and intelligent detection systems capable of responding in real time.

Traditional cybersecurity techniques, while foundational, are often insufficient to address the dynamic and real-time constraints of AVs. A resilience-focused approach—one that utilizes redundancy, diversity, and adaptive response mechanisms—is essential for maintaining critical functionality in the presence of faults or attacks ~\cite{sterbenz2010resilience,ganin2019resilience}. Moreover, the development of lightweight, real-time anomaly detection systems and self-recovery mechanisms are crucial to support operational continuity under adversarial conditions.

This chapter aims to provide a comprehensive overview of the threat landscape in autonomous systems and explore state-of-the-art methodologies for enhancing their security and resilience. We begin by outlining the motivation for AV-specific security measures, followed by a taxonomy of known attacks and an evaluation of current anomaly detection and mitigation techniques. We then introduce a layered threat modeling framework and propose a set of practical defense strategies, supported by experimental demonstrations using a physical AV platform. Finally, we present an AV resilience architecture that integrates detection, validation, and fallback strategies, paving the way for more robust and trustworthy operations of autonomous systems.

\section{Background and Related Work}
\subsection{Major AV components}
Autonomous vehicles are generally organized into three major components: perception, planning, and control \cite{pendleton2017perception,ulbrich2017towards}. Planning and control together define how the vehicle determines a feasible path and executes the necessary maneuvers to follow it safely. Perception, however, forms the foundation of this pipeline, as it equips the vehicle with awareness of its surrounding environment \cite{huang2022multi,wang2019multi}. A perception system integrates data from multiple sensors such as LiDAR, radar, and cameras, enabling the detection, localization, and tracking of objects and road features. The importance of perception lies in its role as the primary source of information for autonomous navigation: without reliable perception, subsequent planning and control processes cannot operate effectively.


\subsubsection{Perception System Attacks}
Recent studies have conducted real-world experiments on remote attacks against these perception sensors \cite{petit2015remote, xu2023sok}. The perception system plays a central role in autonomous driving, as it integrates a diverse suite of sensors such as cameras, LiDAR, radar, ultrasonic sensors, GPS, and Inertial Measurement Unit (IMU) to construct an accurate understanding of the surrounding environment. Since each of these components contributes uniquely to situational awareness, the compromise of even a single sensor can degrade perception accuracy and mislead decision-making modules \cite{ivanov2016attack}.


\subsubsection{Perception Security:}  

To secure vehicle perception, a robust detection system along with a mitigation strategy is crucial. Based on the existing literature (\cite{petit2015remote}, \cite{shin2017illusion, xu2018analyzing, ivanov2016attack, yan2016can, liu2021seeing}), the current approaches for detecting perception sensor attacks against AVs can be categorized as follows: (i) Incorporating redundancy by adding more sensors, (ii) Cross-vehicle verification through Vehicle-to-Vehicle (V2V) communication, and (iii) Relying on alternative sensors and sensor fusion. Although sensor redundancy improves resilience against random attacks, it is ineffective against intentional attacks specific to a particular sensor, where the attacker has prior knowledge of the sensor. Using inter-vehicle communications like V2V to compare sensor measurements is ineffective when the victim vehicle is out of the communication range~\cite{milaat2018decentralized, oligeri2022gps}. Multi-modal fusion, like combining LiDAR and camera data, has shown promising results in improving robustness under benign conditions but still struggles against coordinated multi-sensor adversarial attacks \cite{qi2018frustum, chen2017multi, ku2018joint, tu2021exploring}. 

Recent work has explored adversarial training and certified defense mechanisms for perception models, aiming to improve robustness against crafted perturbations \cite{goodfellow2014explaining,tu2021exploring}. However, these approaches often incur high computational costs and are limited to specific attack models, making them less practical for real-time autonomous driving \cite{zhang2022adversarial, zheng2020efficient}. Thus, there remains a critical need for lightweight, adaptable defenses that can operate under diverse and evolving attack scenarios.

Given these limitations of the existing methods, there is a need for a holistic resilient solution that can comprehensively detect and mitigate attacks without relying on specific sensors, sensor redundancy, or other vehicles and networks.

\section{Redefining the AV Architecture for Threat and Resilience Modeling}
\label{sec:av-architecture}

In this work, the AV stack is intentionally redefined to emphasize resilience and security considerations. While conventional AV architectures typically separate the system into perception, planning, and control, here the stack is organized into distinct functional layers that more directly capture the operational roles and potential threat surfaces relevant to our study. Although the overall layered concept is inspired by prior work~\cite{yousseef2024autonomous}, the specific redefinition presented here is tailored to resilience analysis rather than intended as a universal AV taxonomy.

\begin{itemize}
    \item \textbf{Driver Interface and Control System}: Provides the human driver with inputs and outputs, including steering feedback, dashboard displays, and manual override functions. This layer forms the boundary between human operators and the autonomous driving system. Modern AVs increasingly emphasize intuitive human--machine interfaces (HMIs), ensuring that drivers can safely monitor vehicle status and intervene during edge cases or system failures. 
    
    \item \textbf{Control Layer}: Responsible for low-level actuation such as steering, throttle, and braking, as well as implementing motion control algorithms to execute trajectories generated at higher levels. Advanced architectures often integrate high-frequency feedback loops, redundancy in actuation, and fail-safe control policies to guarantee safety in the presence of disturbances or component failures. 
    
    \item \textbf{Decision-Making Layer}: Manages tactical and strategic planning, including path planning, obstacle avoidance, and maneuver selection. This layer translates high-level goals into executable control commands. Current systems leverage reinforcement learning and optimization-based planners to handle complex traffic scenarios, dynamic obstacles, and rule-based constraints in real time. 
    
    \item \textbf{Knowledge Processing Layer (Data Fusion and Interpretation)}: Integrates heterogeneous data from multiple onboard sensors and external sources to maintain a coherent representation of the environment and vehicle state. In prior studies, this functionality was often included within the Perception Layer, but here it is explicitly separated to emphasize its role in resilience and security. Advanced architectures employ AI-driven sensor fusion and map-matching techniques to reduce uncertainty and dynamically adapt to changing conditions, improving the robustness of situational awareness.

    \item \textbf{Perception Layer}: Processes raw sensor data (camera, LiDAR, radar, GPS, etc.) to perform object detection, tracking, and environmental mapping, which are essential for situational awareness. Recent approaches use deep neural networks for multi-modal perception and adopt adversarially robust methods to reduce susceptibility to spoofing, noise, or environmental disturbances. 
    
    \item \textbf{Communication Framework (V2V, V2I, V2C)}: Supports information exchange between vehicles, infrastructure, and cloud services, enabling cooperative perception, coordinated maneuvers, and extended situational awareness. With the introduction of 5G and Cellular Vehicle to Any, this layer increasingly underpins cooperative driving, low-latency hazard alerts, and real-time map updates. 
    
    \item \textbf{Cross-Layer Coordinator}: Ensures synchronization, resilience, and consistency across all layers of the AV stack, maintaining reliable operations and overall system coherence. This cross-layer is crucial for fault tolerance and safety assurance, as it enables graceful degradation, redundancy management, and system-wide cybersecurity monitoring. 
\end{itemize}

\section{Layer-based Attack Scenarios}
\label{sec:attacks-by-layer}

To systematically evaluate threats in autonomous vehicles (AVs), we adapt the IoT threat modeling framework~\cite{satam2020anomaly} to the AV domain. Each architectural layer is examined in terms of its \textit{core functions}, \textit{attack surfaces}, \textit{impact}, and \textit{mitigation strategy}. This structured approach enables a consistent mapping between functional responsibilities and their associated vulnerabilities. Representative cases derived from this methodology are presented in Table~\ref{tab:attack-mapping}.

\subsection{Layer Descriptions}

The AV architecture can be conceptualized as a hierarchy of interdependent functional layers, each responsible for specific operations and each presenting unique attack surfaces.  

At the top, the \textbf{Driver Interface and Control System} links the human operator to the vehicle via dashboards, controls, and manual override mechanisms. Compromise at this layer may confuse the driver, override legitimate commands, or inject malicious control inputs. Beneath it, the \textbf{Control Layer} governs low-level actuation, including throttle, steering, and braking. Attacks here can silently alter vehicle dynamics, producing deviations that are difficult to detect but potentially dangerous.  

Building upward in abstraction, the \textbf{Decision-Making Layer} manages tactical and strategic behaviors such as lane changes, obstacle avoidance, and path planning. Manipulating decision models or outputs may lead to unsafe maneuvers or traffic violations. The \textbf{Knowledge Processing Layer}, which fuses multisensor and external data into a coherent world model is needed to support the decision-making processes. Poisoned streams or corrupted fusion algorithms can distort situational awareness and undermine downstream decisions.  

The \textbf{Perception Layer} translates raw inputs from LiDAR, radar, cameras, and GPS into object detections, classifications, and environmental maps. It is especially exposed to spoofing, adversarial perturbations, and physical-world manipulations (e.g., altered traffic signs). Complementing perception, the \textbf{Communication Framework} handles information exchange with  other vehicles, infrastructure, and the cloud via V2V, V2I, and V2C links, respectively. Replay, spoofing, or denial-of-service attacks on this layer can disrupt cooperative driving and erode trust.  

Finally, the \textbf{Cross-Layer Coordinator} ensures synchronization and resilience across all layers. Because it influences the entire stack, faults or injected errors here can cascade through the architecture, destabilizing the system as a whole.  

Together, these layers provide a structured visibility through which attacks can be mapped, analyzed, and countered using the five-step methodology.

\textbf{In this chapter, we focus primarily on the Perception and Knowledge Processing layers as case studies to demonstrate how resilience and security measures can be integrated into an AV system.}

\begin{table}[!htbp]
\centering
\resizebox{\textwidth}{!}{
\begin{tabular}{|p{3cm}|p{4cm}|p{5cm}|p{5cm}|}
\hline
\textbf{Layer} & \textbf{Attack Surface (AS)} & \textbf{Impact} & \textbf{Mitigation Strategy} \\ \hline

\textbf{Driver Interface and Control System} & Malicious driver input injection; compromised dashboard/override signals& Unauthorized override of system behavior; driver confusion leading to unsafe maneuvers  & Authenticate commands, secure UI design, redundant safety checks for manual override \\ \hline

\textbf{Control Layer} & Command injection in steering/braking; kernel-level rootkits & Silent manipulation of vehicle dynamics, risk of collisions & Runtime integrity monitoring, secure boot, anomaly detection in actuation signals \\ \hline

\textbf{Decision-Making Layer} & Tampered ML models; adversarial inputs to planning module & Unsafe maneuvers, rule violations, unsafe path selection & Model validation, adversarial robustness testing, consensus-based decision redundancy \\ \hline

\textbf{Knowledge Processing Layer} & Data poisoning; sensor fusion manipulation & Distorted environmental model, leading to misinformed decisions & Trusted data provenance, anomaly detection in fused data, hash-based detection, back up and fall back \\ \hline

\textbf{Perception Layer} & Sensor spoofing (LiDAR, radar, camera); adversarial image perturbations & Incorrect object detection, phantom obstacles, missed hazards & Sensor redundancy, adversarial training, anomaly detection \\ \hline

\textbf{Communication Framework} & V2V/V2I spoofing, replay, DoS, packet manipulation & Loss of situational awareness, degraded coordination between vehicles & Encryption, authentication protocols, rate limiting, intrusion detection systems \\ \hline

\textbf{Cross-Layer Coordinator} & Early-stage poisoning cascading through layers & System-wide compromise, cascading unsafe behavior & Cross-layer consistency checks, layered anomaly detection, resilient failover strategies \\ \hline

\end{tabular}
}
\caption{Threat Modeling for the proposed AV architecture. Each row summarizes the attack surface, safety impact, and candidate mitigation strategy.}

\label{tab:attack-mapping}
\end{table}

\section{Security and Resilience Methodology}

\begin{figure}[h]
    \centering
    \includegraphics[width=\linewidth]{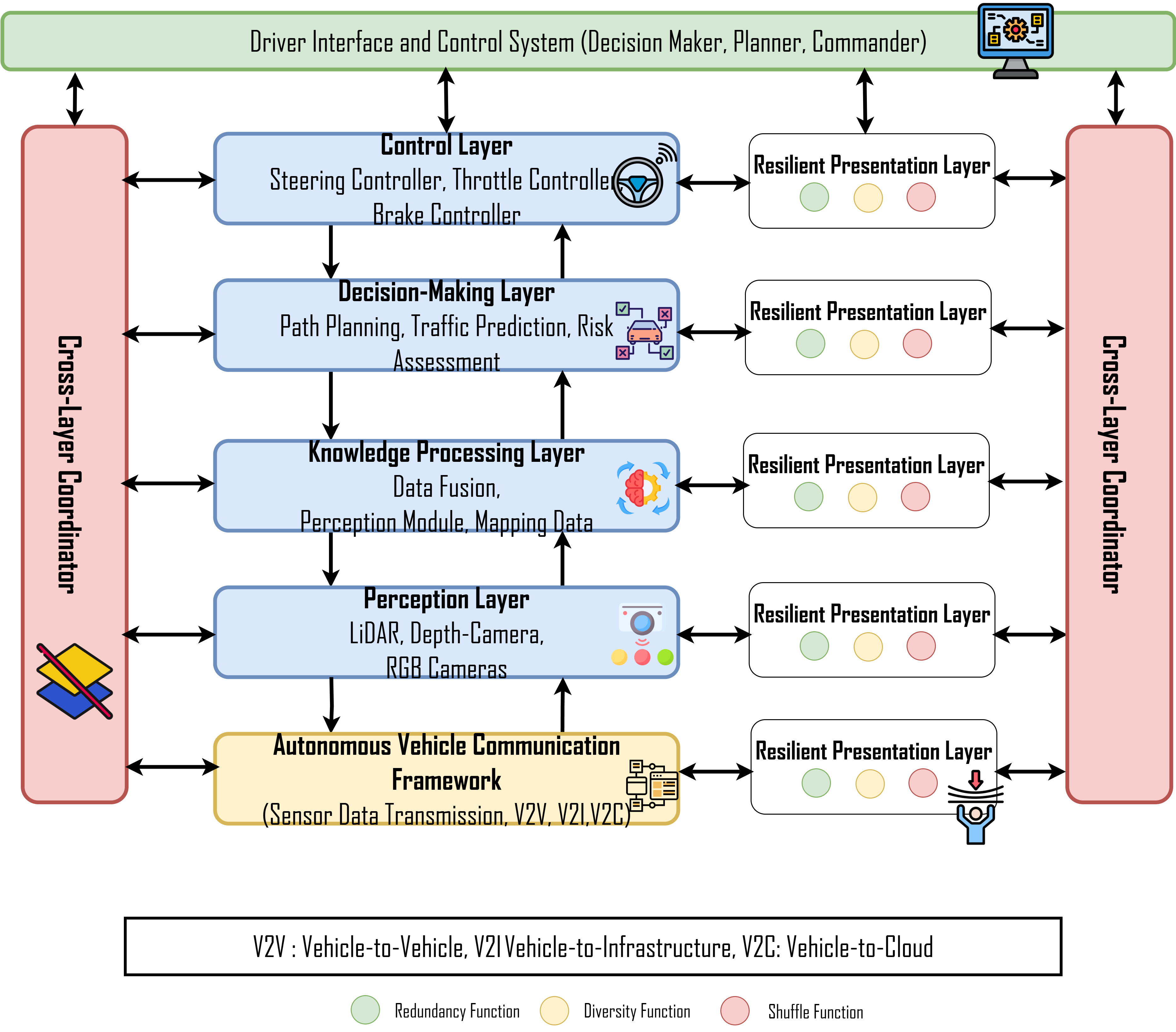}
    \caption{AVR architecture integrated across AV functional layers and coordinated by a Cross-Layer Coordinator.}

    \label{fig:resilient_architecture}
\end{figure}

Based on the system architecture and identified mitigation strategies, we present an Autonomous Vehicle  Resilient (AVR) Architecture in conjunction with targeted threat and intrusion detection techniques to enhance the security and operational robustness of autonomous vehicles. The overall concept of the AVR architecture is illustrated in Figure~\ref{fig:resilient_architecture}, while the anomaly detection framework is depicted in Figure~\ref{fig:detection_flow}. Detailed mechanisms are discussed in further detail in the following section. \label{sec:methods}

\subsection{AVR Architecture}
The \textbf{AVR architecture} is designed to operate across all functional layers of the autonomous vehicle stack, providing a systematic framework to anticipate, detect, and recover from failures or attacks. Its resilience is achieved through three complementary functions:

\begin{itemize}
    \item \textbf{Redundant Functions} – critical modules are backed up by verified functionally equivalent redundant modules to ensure continuity of operations in the event of component failures or malicious tampering. This approach reduces the risk of single points of failure and provides a safety net for vital system operations. 
    
    \item \textbf{Diversity Functions} – by incorporating heterogeneous sensors, algorithms, and models, the system mitigates common-mode failures and increases robustness against attacks that depend on existing vulnerabilities in a single type of module or data source.
    
    \item \textbf{Shuffle Functions} – periodic reconfiguration of software, communication channels, and system parameters introduces a high degree of ambiguity and consequently reduces the likelihood that adversaries can exploit deterministic system behaviors.
\end{itemize}

A dedicated \textbf{Cross-Layer Coordinator} oversees these functions by monitoring anomalies, enforcing consistency across layers, and triggering dynamic reconfiguration when anomalies are detected. This ensures that multi-stage or stealthy attacks are isolated and mitigated before they can propagate, effectively enhancing overall AV system resilience.

\subsubsection{Backup and Fallback Mechanism}

We illustrate the backup and fallback concept through two operational scenarios: a \textit{normal state} and a \textit{failure state}. In the normal state,  the \textbf{Primary module ($P$)} executes normally, and no intervention is required. The system continues to operate smoothly as long as the detection function $D(\cdot)$ reports no anomalies. 

In contrast, the failure state occurs when $D(P)=1$ signals a compromise or fault in the primary module. At this point, control is immediately transferred to the \textbf{Fallback module ($F$)}, which is a verified and trusted redundant module. The transition ensures that the system preserves safety-critical functions even under attack or malfunction. 

\begin{figure}[h]
    \centering
    \includegraphics[width=0.7\textwidth]{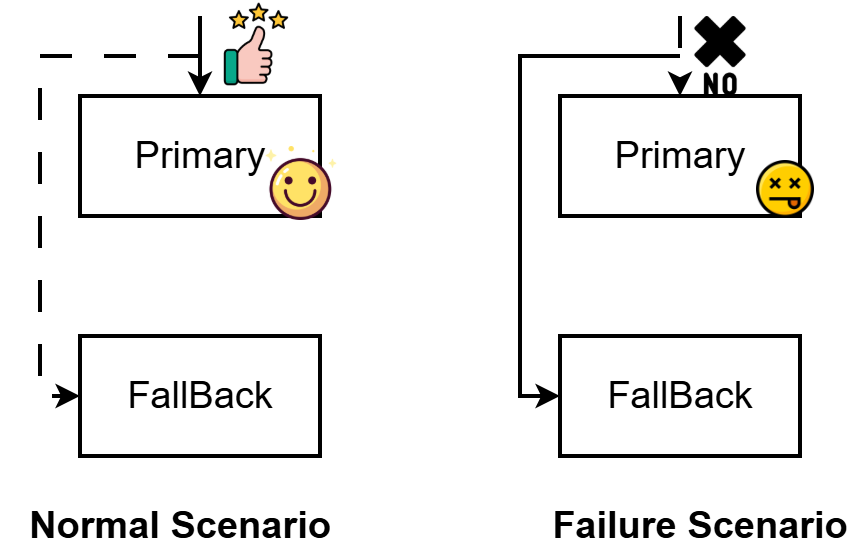}
    \caption{Two operational scenarios for the backup/fallback mechanism: (1) happy state with primary execution, (2) failure state with automatic switchover to verified fallback.}
    \label{fig:fallback}
\end{figure}

\begin{figure}[htbp]
    \centering
    \includegraphics[width=0.6\linewidth]{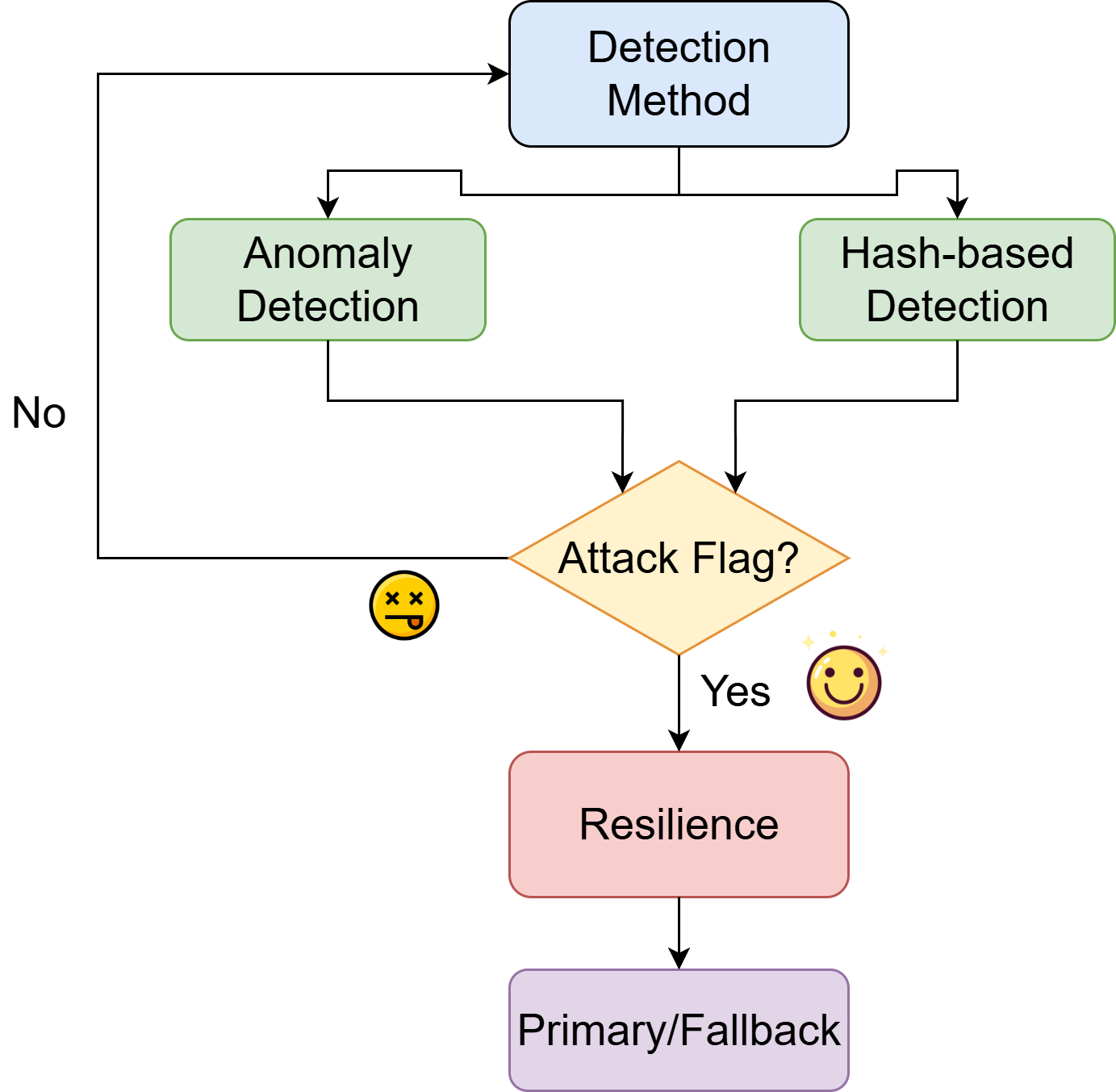}
    \caption{Intrusion Detection Techniques: anomaly-based (perception) and hash-based (software/system integrity).}
    \label{fig:detection_flow}
\end{figure}

Figure~\ref{fig:fallback}  depicts this mechanism: under normal conditions the primary operates, but upon detection of failure, the fallback takes over seamlessly. This dual-scenario design embodies the redundancy principle of the AVR architecture shown in Figure 1.

\subsection{Intrusion Detection Techniques}
An intrusion detection system (IDS) is used in this work to provide real-time indications of compromises and to trigger resilience responses. We employ two complementary techniques: (i) anomaly-based detection to identify deviations in runtime behavior at the perception level, and (ii) hash-based integrity validation to detect unauthorized modifications to software artifacts (e.g., model weights and configuration files). These are described in the following subsections and together provide both behavioral and software integrity assurance.

\subsubsection{Anomaly-Based Detection}




Anomaly-based detection techniques identify unusual or unexpected system behavior that may indicate a cyberattack. It relies on a normal behavior model of the whole system or system components. System activities are continuously monitored and compared against the baseline normal behavior model. When instantaneous behavior deviates beyond a predefined normality threshold, it is flagged as anomalous, which may indicate a potential intrusion.

In the context of AVs, anomaly-based intrusion detection depends on accurately characterizing the vehicle’s normal operational behavior, which is typically learned through offline modeling and training. If an attacker attacks vehicle components (for example, by injecting malicious data into sensors) that cause the vehicle to behave abnormally, the IDS identifies this deviation and raises an alert. A key strength of anomaly-based IDS is its capability to identify previously unseen attacks better than other detection algorithms; however, developing an accurate normal behavior model is challenging. Inaccuracies in this model can lead to high false positive or false negative errors.

\begin{figure}[h]
\centerline{\includegraphics[width=12cm]{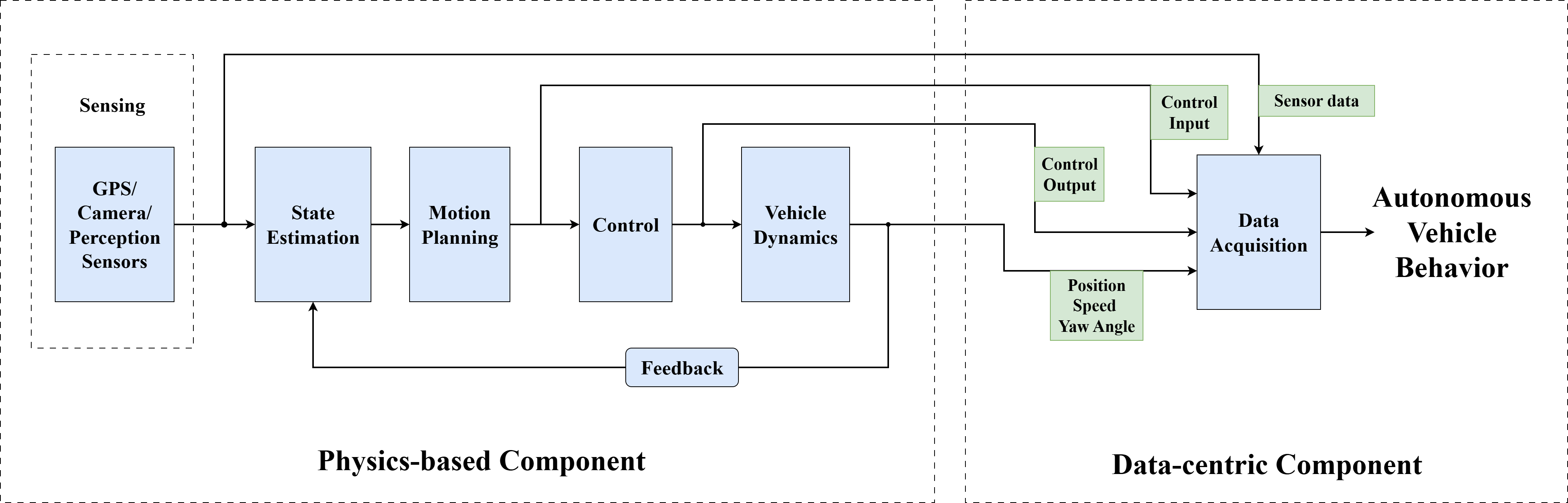}}
\caption{Simplified Autonomous Vehicle Behavior Model} \label{avbm_fig}
\end{figure}

To address this, we develop the normal behavior model of AVs using a combination of a mathematical model of the vehicle system called the Autonomous Vehicle Behavior Model (AVBM) and machine learning algorithms. The mathematical model characterizes the instantaneous state of the vehicle, while the learning-based components account for the data-driven variations in normal operation. A simplified overview of the AVBM is illustrated in Fig.~\ref{avbm_fig}, and further details are available in \cite{abrar2023anomaly, abrar2024gpsids}.

\paragraph{Anomaly-based Intrusion Detection Methodology.}

\begin{figure}[h]
\centerline{\includegraphics[width=7cm]{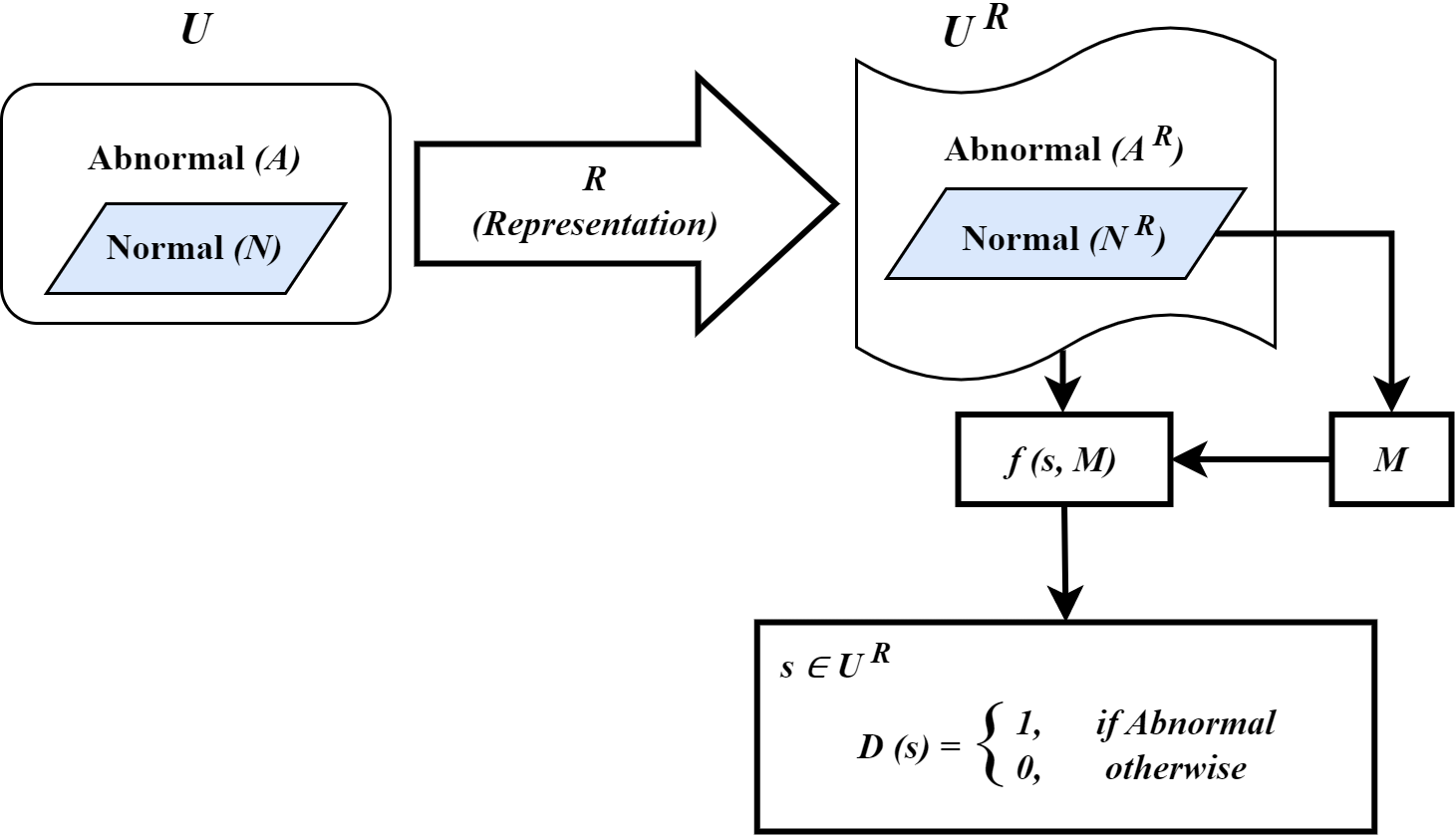}}
\caption{Anomaly-based Intrusion Detection Methodology} \label{ABAfig}
\end{figure}


Fig. \ref{ABAfig} shows a diagram of anomaly-based intrusion detection technique for AVs. We define the anomaly detection approach over a finite set of driving events $U$. Set $U$ is divided into two subsets: $Normal$ driving events $N$ and $Abnormal$ driving events $A$, such that: $N \cup A = U$ and $N \cap A = \emptyset$. A representation map $R$ is used to model $U$, which maps driving events in $U$ to patterns in $U^R$. Similarly, $N$ and $A$ driving sets are mapped to $N^R$ and $A^R$ using the representation map $R$. These can be mathematically expressed as: $U \xrightarrow{R} U^R$, $N \xrightarrow{R} N^R$, $A \xrightarrow{R} A^R$, and $N^R \cup A^R = U^R$. A detector $D$ is defined as $D = (f, M)$; where $f$ denotes the normal behavior characterization function expressed as $f: U^R \times M \Rightarrow [0, 1]$, and $M$ is denoted as the system memory that stores the normal behavior model extracted from the set of normal driving events $N^R$. Function $f$ specifies the degree of abnormality of a sample $s \in U^R$ by comparing it with $M$. The higher the value of $f(s, M)$, the more abnormal the sample is. If the value of $f(s, M)$ exceeds a predefined threshold $\mathbb{T}$, detector $D$ raises an alarm that indicates the occurrence of an abnormal event. Detector $D$ can be mathematically expressed as follows \cite{abrar2023anomaly}:

\begin{equation*}
D(s) = 
\begin{cases}
    Abnormal \hspace{1cm} if \quad f(s, M) > \mathbb{T}\\
    Normal \hspace{1.3cm} otherwise
\end{cases}
\end{equation*}

\subsubsection{Hash-Based Integrity Validation}
Hash-based integrity validation is used in this work as a lightweight, runtime safeguard for the QCar perception stack \cite{uriawan2024authenticate} \cite{TsaiHariri2025CSCloud}. Specifically, the electronic control unit (ECU) periodically computes a cryptographic hash (SHA-256) over a predefined set of integrity-critical artifacts, including the deployed YOLOv8 model weights and associated configuration files (e.g., detection thresholds and class-to-label mappings). The computed digest is compared against a trusted baseline hash stored in a protected location.

A hash mismatch is treated as a software-tampering event. Upon detection, the system logs the anomaly (timestamped) and immediately triggers the resilience response: execution is switched to a pre-verified fallback/backup perception module and trusted files are restored, allowing the vehicle to maintain safety-critical behaviors without requiring a system restart. The fallback remains active in an event-driven manner as long as the integrity violation persists, and normal operation is automatically restored once the expected hash values are re-established.

\section{Experimental Setup and Results}
\label{sec:experiments}
Experiments were conducted using Quanser's QCar self-driving platform, a small-scale, high-fidelity testbed for perception, planning, and control research \cite{qcar2}. The vehicle is equipped with a drive motor, steering servo, 4 cameras, a 2D LiDAR, an Intel RealSense depth camera, and a 9-axis inertial measurement unit (IMU). An onboard NVIDIA Jetson Orin processor executes control algorithms, receives commands via Wi-Fi from a ground station, and transmits sensor data back for monitoring.

As mentioned in previous sections, our experiments focus on attacks targeting the Perception and Knowledge Processing layers. Perception-layer attacks include sensor-level manipulations, such as laser blinding of the depth camera, degrading obstacle detection and braking. Knowledge-layer attacks involve software manipulations, including modifications to the YOLOv8 model and configuration files, aiming to compromise data fusion and world-model integrity. These experiments evaluate the effectiveness of resilience mechanisms, including anomaly detection, hash-based integrity validation, and backup/fallback strategies.

Two experimental setups were employed, each corresponding to a distinct threat detection technique and backup and fallback for resilience.

\subsection{Anomaly-based Intrusion Detection Setup}

\subsubsection{Attack Experiment}

\begin{figure}[h]
\centerline{\includegraphics[width=11cm]{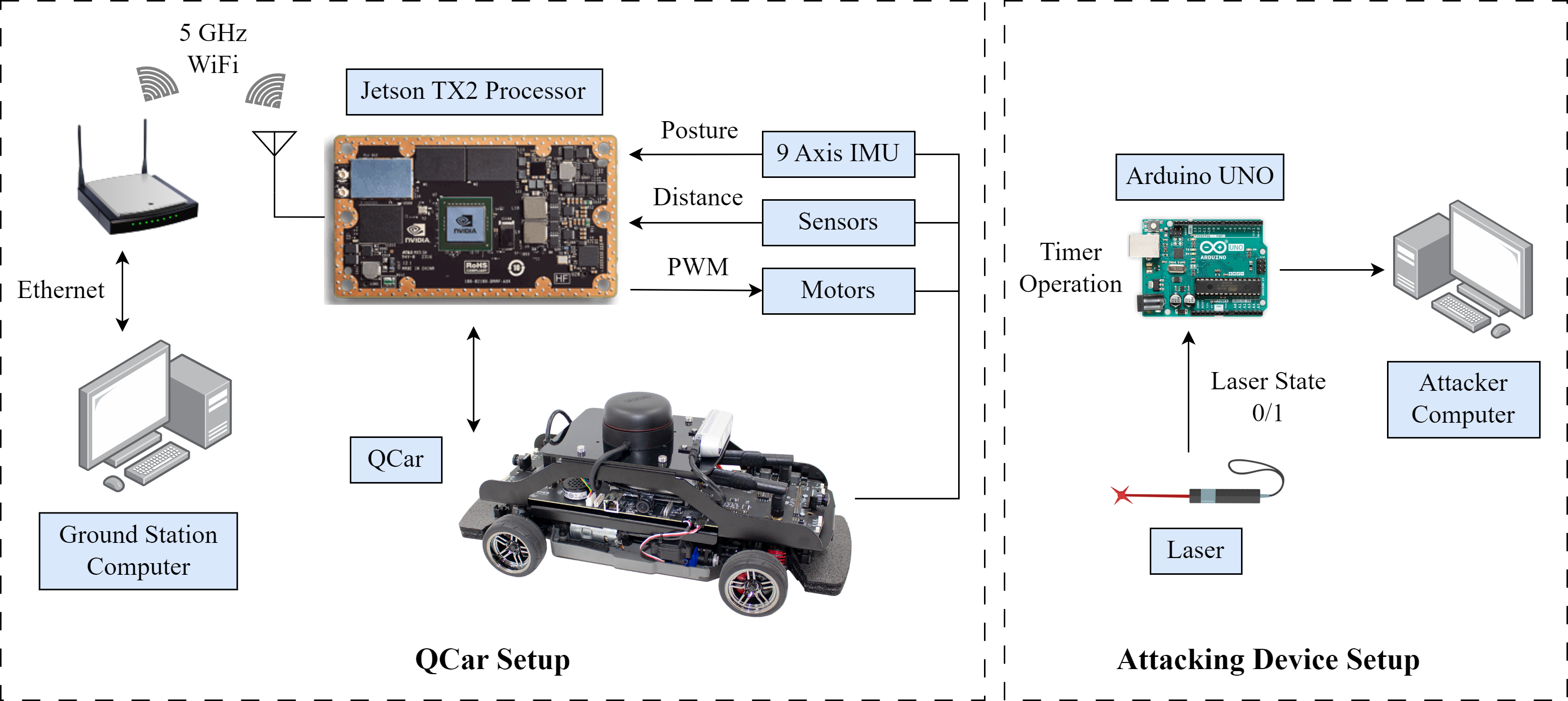}} 
\caption{Setup of the depth camera blinding attack experiment on QCar}
\label{qcar setup}
\end{figure}

In this experimental setup, we performed a real-world depth camera blinding attack against the onboard depth camera of QCar to degrade obstacle detection and braking capabilities during autonomous driving. A laser blinding attack was performed as explained by Yan et al. \cite{yan2016can}. The hardware setup of the attack device consisted of a wide-beam laser controlled by an Arduino, which recorded digital clock timestamps corresponding to the laser’s on and off states. The device generated a binary signal, where a value of 1 denoted laser activation and 0 denoted deactivation. Both the QCar and the attack device operated at the same data sampling rate and were initialized simultaneously to ensure temporal alignment. Fig. \ref{qcar setup} illustrates the setup used for the anomaly-based intrusion detection experiments.

\subsubsection{Data Collection: AVP-Dataset}

The Autonomous Vehicle Behavior Model in Fig. \ref{avbm_fig} highlights the essential parameters necessary for estimating the present and future states of the vehicle. This helps model the normal behavior of the vehicle and guides us in collecting data of specific features for our experimental analysis. The Autonomous Vehicle Perception Dataset (AVP-Dataset) is prepared by incorporating the parameters obtained from the AVBM. The full AVP-Dataset is publicly available at \cite{mehrababrar_AVPDataset, abrar2023anomaly}.


\subsection{Hash-based Integrity Validation Setup}

\subsubsection{Attack Experiment}

In this experimental setup, we conducted a \emph{software-based perception tampering attack} on the QCar’s onboard NVIDIA Jetson AGX Orin main ECU to evaluate the effectiveness of hash-based integrity validation. The attack degrades obstacle detection and braking behavior by modifying internal perception parameters at runtime, such as YOLOv8 model configuration files, object detection thresholds, or class-to-label mappings. Unlike physical-layer attacks, this approach does not interfere with sensors directly and instead represents realistic advanced driver-assistance systems (ADAS) and autonomous vehicle threats originating from within the ECU.

To assess robustness and repeatability, five independent software tampering events were triggered under each experimental condition, with each event treated as an individual attack trigger and distributed across multiple autonomous driving runs. No explicit upper limit was imposed on the number of triggers, allowing the system to handle repeated tampering events without requiring a restart or manual intervention.

When an integrity violation is detected, the system does not issue an emergency stop. Instead, it automatically switches to a pre-verified fallback perception module, enabling the vehicle to continue operating while preserving safety-critical functions. The fallback mechanism remains active as long as the abnormal configuration persists and is released once the trusted state is restored. As a result, the vehicle does not unconditionally stop after an attack but continues to drive safely and executes appropriate behaviors, such as stopping for pedestrians or stop signs, based on the fallback perception output.

\begin{figure}[h]
  \centering
  \includegraphics[width=0.6\textwidth]{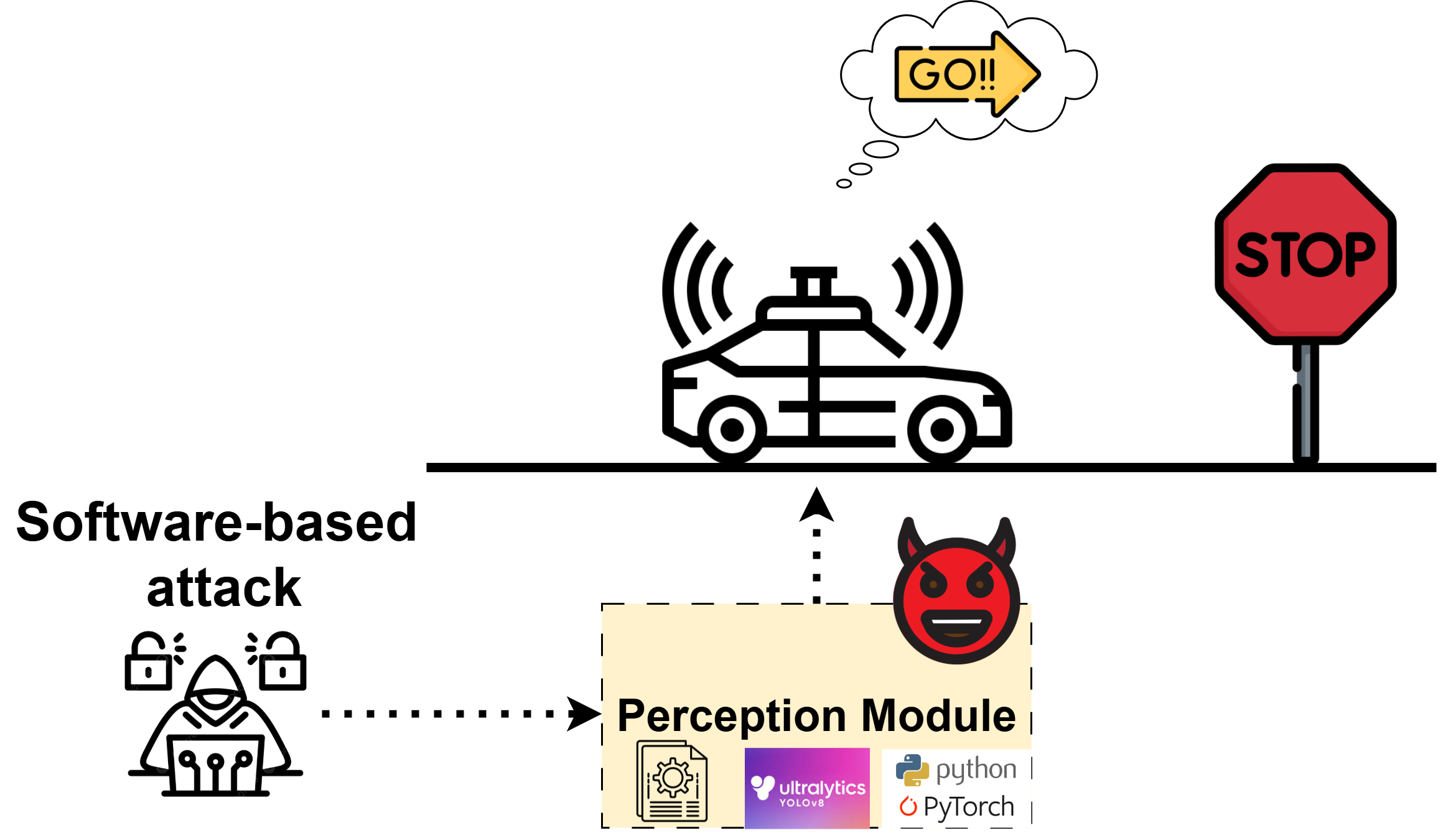}
  \caption{Software-based attack: ECU perception module compromise on QCar2.}
  \label{fig:attack_scenario}
\end{figure}

\subsubsection{Data Collection: Integrity-Validation Dataset}

Data was collected under normal operation and during software attacks. In the normal scenario, the YOLOv8 model and configuration files remained unaltered, with hashes matching the trusted baseline value. During attacks, modified models or configuration files generated hash mismatches, which were logged by the system. The Integrity-Validation Dataset contains synchronized traces of both normal and attack scenarios, capturing system responses, detection outcomes, and integrity validation results. 

\subsection{Experimental Results}
\subsubsection{Results: Anomaly-Based Detection}

This experiment evaluates the performance of seven supervised binary classification models, namely Random Forest (RF),  Logistic Regression (LR), XGBoost (XGB), K-Nearest Neighbor (KNN), Support Vector Classifier (SVC), Multi-Layer Perceptron (MLP), and Naive Bayes (NB), using the AVP-Dataset. The evaluation was conducted on the full dataset comprising 88{,}976 samples, including 45{,}001 normal instances (50.57\%) and 43{,}975 abnormal instances (49.43\%). Classifier performance was assessed using four standard metrics: accuracy, precision, recall, and F1 score. Since the primary objective is to identify abnormal samples, all metrics were computed by treating the abnormal class as the positive class. To ensure unbiased evaluation, a stratified 5-fold cross-validation scheme was employed to preserve class balance across folds. In each fold, 80\% of the data were used for training and the remaining 20\% were used for testing.

The comparative performance of the evaluated models is summarized in Table~\ref{mlperformance}. The results indicate that \textbf{Random Forest} achieved the highest accuracy, precision, recall, and F1 score among all classifiers. Logistic Regression, K-Nearest Neighbor, and Multi-Layer Perceptron demonstrated competitive performance with comparable F1 scores, following closely behind Random Forest.

\begin{table}[h]
\caption{Depth Camera Blinding Attack Detection Performance of Different Machine Learning Models}\label{mlperformance}
\centering
\begin{tabular}{|c|c|c|c|c|c|c|c|}
\hline
\textbf{Metrics} & \textbf{RF} & \textbf{LR} & \textbf{XGB} & \textbf{KNN} & \textbf{SVC} & \textbf{MLP} & \textbf{NB} \\
\hline
Precision & \textbf{0.801} & 0.800 & 0.751 & 0.793 & 0.742 & 0.786 & 0.692 \\
Recall   & \textbf{0.894} & 0.872 & 0.876 & 0.877 & 0.913 & 0.874 & 0.923 \\
F1 Score & \textbf{0.845} & 0.834 & 0.809 & 0.833 & 0.819 & 0.828 & 0.791 \\
Accuracy & \textbf{0.838} & 0.829 & 0.795 & 0.826 & 0.800 & 0.820 & 0.759 \\
\hline
\end{tabular}
\end{table}


\subsubsection{Tuning the Detection Threshold $\mathbb{T}$ for Anomaly-based Detection}

\begin{figure}[h]
\centerline{\includegraphics[width=7.5cm]{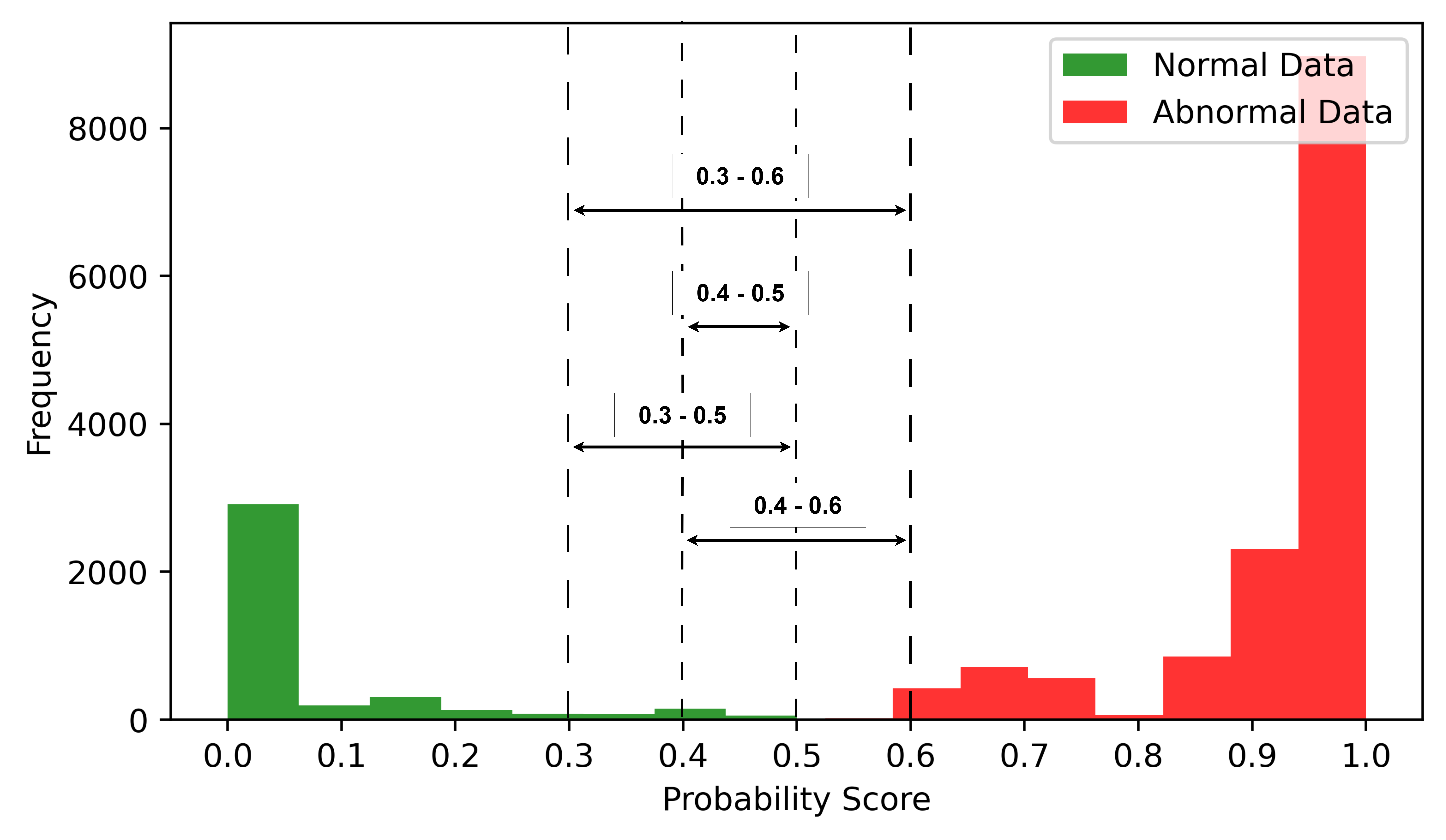}} 
\caption{Distribution of Random Forest abnormal-class probability scores for normal and attack data. The evaluation fold consists of 3{,}893 normal samples and 13{,}902 attack samples. An effective detection threshold lies in the range $\mathbb{T}=0.4$--$0.5$.}

\label{margin}
\end{figure}


In this experiment, we investigated the selection of an appropriate predefined threshold $\mathbb{T}$ for the Anomaly Behavior Analysis framework by examining the predicted probability scores associated with normal and abnormal samples. Since the \textbf{Random Forest} model exhibited the strongest overall performance in Experiment~2, it was selected for this experiment. The probability outputs were visualized to identify a suitable separation margin for threshold placement. Fig.~\ref{margin} illustrates the overlaid probability score distributions for normal and abnormal data, where a clear distinction between the two classes can be observed, as highlighted by the dotted boundaries.

Table~\ref{margin table} reports the corresponding misclassification rates as we widen the decision margin. The results indicate that choosing the threshold $\mathbb{T}$ within the interval of \textbf{0.4 to 0.5} provides the most favorable trade-off. In this range, 155 normal samples are incorrectly labeled as abnormal, while no attack samples are misclassified, resulting a False Positive Rate of 0.0398 (3.98\%) and a False Negative Rate of 0. Thus, setting the predefined threshold $\mathbb{T}$ to a value between 0.4 and 0.5 for the Anomaly Behavior Analysis approach ensures an optimal attack detection capability.

\begin{table}[h]
\caption{False Positive (FP) and False Negative (FN) Rates for Different Detection Margins}
\label{margin table}
\centering
\begin{tabular}{|c|c|c|c|c|}
\hline
\multirow{2}{*}{\begin{tabular}[c]{@{}c@{}}Detection\\ Margin\end{tabular}} &
\multirow{2}{*}{\begin{tabular}[c]{@{}c@{}}Normal Data\\ Misclassified\end{tabular}} &
\multirow{2}{*}{\begin{tabular}[c]{@{}c@{}}Attack Data\\ Misclassified\end{tabular}} &
\multirow{2}{*}{\begin{tabular}[c]{@{}c@{}}FP\\ Rate\end{tabular}} &
\multirow{2}{*}{\begin{tabular}[c]{@{}c@{}}FN\\ Rate\end{tabular}} \\
& & & & \\
\hline
\textbf{0.4 -- 0.5} & \textbf{155} & \textbf{0}  & \textbf{0.0398} & \textbf{0} \\
0.3 -- 0.5          & 293          & 0           & 0.0752         & 0 \\
0.4 -- 0.6          & 155          & 26          & 0.0398         & 0.00187 \\
0.3 -- 0.6          & 293          & 26          & 0.0752         & 0.00187 \\
\hline
\end{tabular}
\end{table}


\subsubsection{Results: Hash-Based Integrity Validation}

\begin{figure}[ht]
    \centering
    \includegraphics[width=0.7\linewidth]{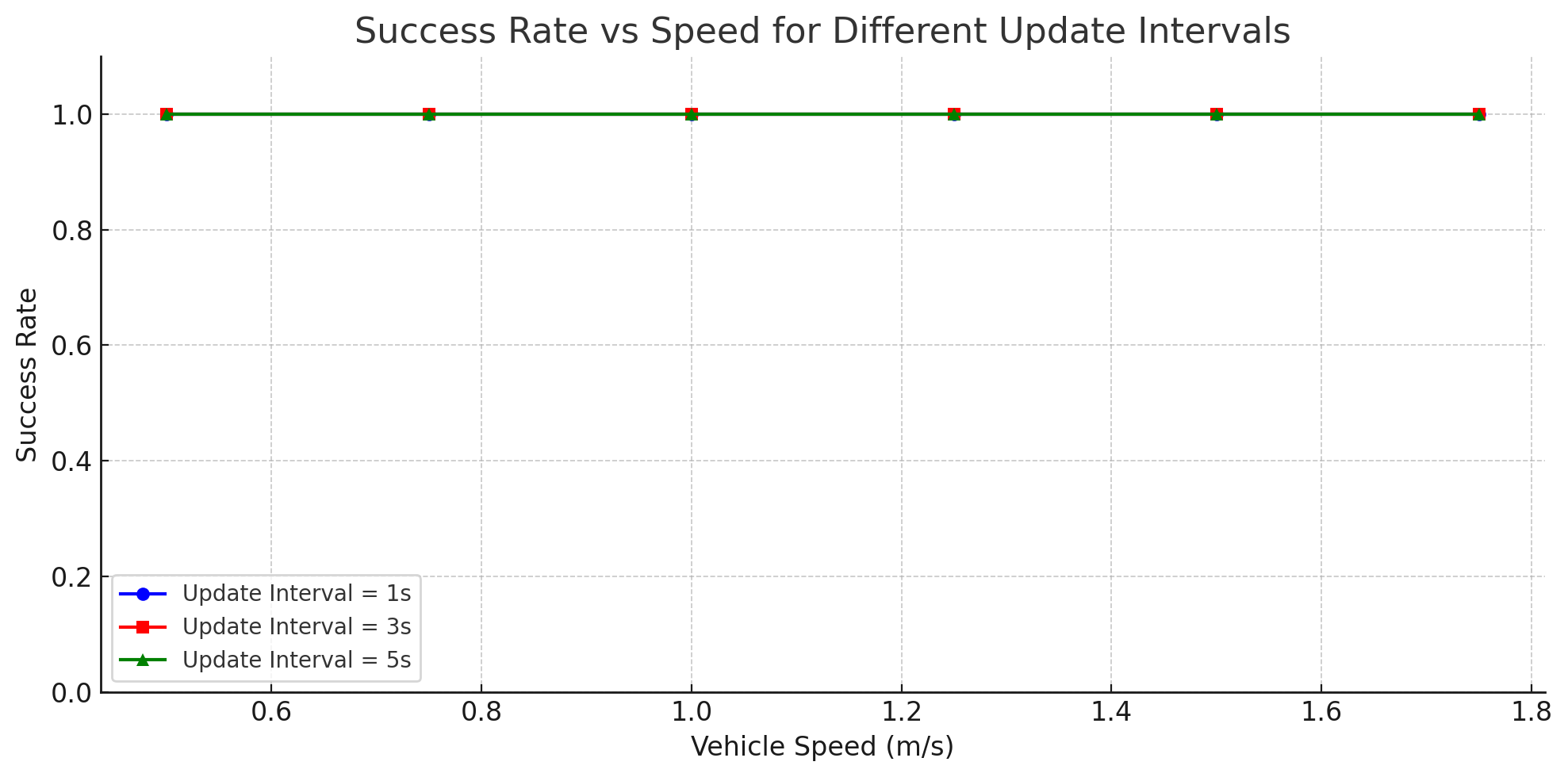}
    \caption{Success rate of hash-based integrity validation across vehicle speeds (0.5--1.75~m/s) and hash update intervals (1~s, 3~s, and 5~s), showing reliable detection of all software modifications.}
    \label{fig:success_rate}
\end{figure}

\paragraph{Detection Accuracy and Timing.}
We evaluated the hash-based detection mechanism on QCar under varying speeds from \textbf{0.5~m/s} to \textbf{1.75~m/s} with update intervals of \textbf{1 s}, \textbf{3 s}, and \textbf{5 s}. Across all configurations, the system achieved a \textbf{100\% success rate}, demonstrating that hash-based validation reliably detects any modification to the model or configuration regardless of vehicle speed or update frequency (Figure~\ref{fig:success_rate}).  

\paragraph{Scalability Discussion.}
In a full-scale vehicle with dozens of concurrent software modules, continuous hashing of all artifacts at high frequency may be impractical. In practice, integrity validation can be scoped to safety-critical artifacts only and executed using staggered schedules or event-driven triggers (e.g., on update, on process restart, or on policy-defined checkpoints), which reduces CPU and I/O contention while preserving tamper detection capability.

\paragraph{Resilience Results.}

\begin{figure}[ht]
    \centering
    \includegraphics[width=0.9\linewidth]{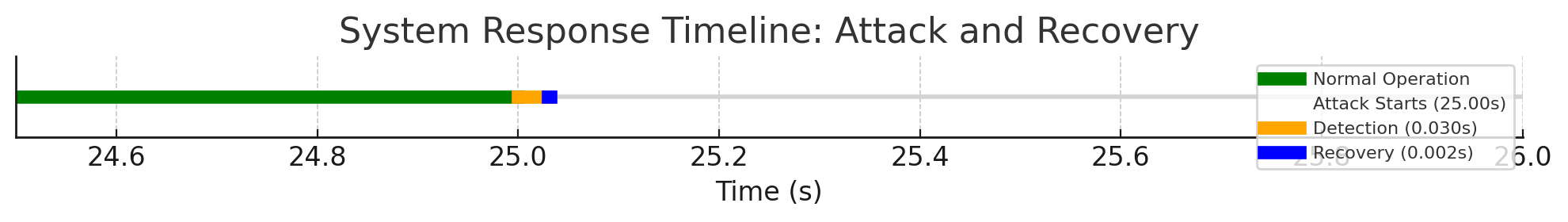}
    \caption{System response timeline under software tampering. The attack is detected via hash mismatch within 0.030~s after onset, and the trusted configuration is restored within 0.002~s, preventing unsafe vehicle behavior.}
    \label{fig:response_timeline}
\end{figure}

\paragraph{System Resilience Evaluation.}
Beyond detection, we measured the system’s response to attacks, focusing on how backup and fallback mechanisms maintain safe operation. Using hash-based monitoring, the system detected a malicious update within \textbf{0.030 s} after its onset at \textbf{25.00 s}. The built-in backup mechanisms immediately restored the trusted configuration and model in \textbf{0.002 s} (Figure~\ref{fig:response_timeline}), preventing any unsafe behavior. 

These results emphasize the importance of having robust fallback strategies: in “normal” scenarios where no attack occurs, the backup remains idle but ready, ensuring normal operation is unaffected; in “attack (abnormal)” scenarios with tampering, the fallback enables the system to revert to a safe, trusted state before any compromised module can influence vehicle behavior. This demonstrates that the combination of fast detection and reliable backup is critical for maintaining both safety and operational continuity in autonomous vehicles.

\begin{figure}[ht]
    \centering
    \includegraphics[width=0.6\linewidth]{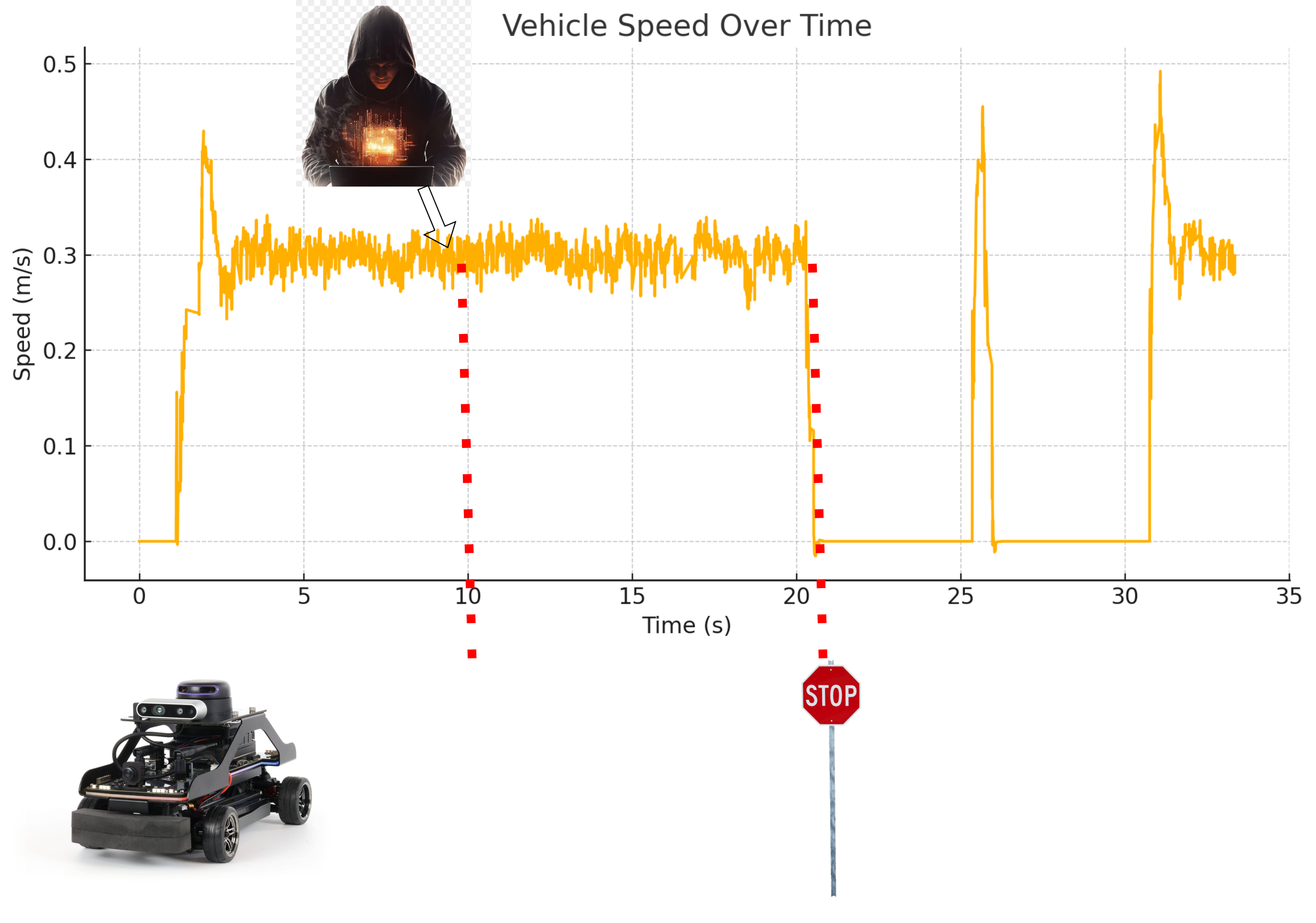}
\caption{QCar speed profile during ECU tampering. After the attack is injected at 10~s, the vehicle continues operating under the fallback perception module and correctly stops at a STOP sign at 20~s, demonstrating preserved safety behavior without an emergency halt.}

    \label{fig:physical_response}
\end{figure}

\paragraph{Physical Validation and Behavior Analysis.}
We further validated the proposed resilience mechanisms on the QCar testbed. During normal operation, the vehicle speed stabilized near \textbf{0.33~m/s}, corresponding to low-speed operation on the scaled platform and being qualitatively comparable to approximately \textbf{12~km/h} driving in a full-scale vehicle. At \textbf{10~s}, a software tampering attack was introduced; despite this, the vehicle correctly halted at a \textbf{STOP} sign at \textbf{20~s}, illustrating that the backup and recovery mechanisms successfully maintained functional safety (Figure~\ref{fig:physical_response}).

This physical experiment demonstrates the practical value of the resilience layer: it not only detects attacks but also ensures that safety-critical behaviors are preserved, preventing unsafe outcomes even under real-world attack conditions.  

\begin{backmatter}

\section{Conclusion}
As autonomous vehicles (AVs) become integral to the future of transportation systems, it becomes critically important to enhance security and resilience of AVs. This chapter has addressed the multifaceted vulnerabilities that AVs face, stemming from their dependency on complex sensor systems, wireless networks, and decision-making processes. By categorizing potential threats across various architectural layers, we have highlighted several ways in which these vehicles can be compromised, how they can be detected and mitigated. 

The introduction of an AV Resilient architecture, characterized by redundancy, diversity, and adaptive reconfiguration, offers a robust framework to tolerate AV attacks. The integration of innovative intrusion detection techniques, such as anomaly detection and hash-based verification, further enhances the security and protection of the AV operations. The successful experimental validation conducted on the Quanser QCar platform exemplifies the practical effectiveness of these strategies in real-world scenarios, particularly in thwarting depth camera blinding attacks and software tampering. 

Ultimately, this work establishes a critical link between threat modeling and the implementation of practical defense mechanisms, paving the way for more resilient autonomous vehicle systems. As we advance toward a future where AVs play an essential role in our transportation landscape, it is imperative that we continue to refine and evolve these security and resilience measures. By prioritizing resilience, we not only enhance the safety and trustworthiness of autonomous vehicles but also contribute to the broader goal of creating intelligent transportation systems that can adapt to ever-evolving challenges in our complex digital world. 
\FloatBarrier  

\section*{Conflict of interest}
The author declares no conflict of interest.

\section*{Acknowledgments}

This work was supported in part by the National Science Foundation (NSF) under Grant 1624668, Grant 1921485, Grant 1303362 (Scholarship-for-Service), Grant 2413009, and the WISPER Center. Support was also provided by the 2025 Technology and Research Initiative Fund/National Security Systems Initiative administered by the University of Arizona, Office of Research and Partnerships, funded under Proposition 301, the Arizona Sales Tax for Education Act (2000). Any opinions, findings, conclusions, or recommendations expressed in this chapter are those of the author(s) and do not necessarily reflect the views of the NSF.
Portions of this manuscript were linguistically polished with the assistance of AI-based language tools (ChatGPT) for grammar and clarity. The authors are solely responsible for the content and technical accuracy of the work. Portions of this chapter are based on previously published work that appeared as a preprint under the title
“GPS-IDS: An Anomaly-Based GPS Spoofing Attack Detection Framework for Autonomous Vehicles,”
available on arXiv (arXiv:2405.08359), 2024.

\end{backmatter}


\end{document}